\documentclass[10pt, conference, compsocconf]{IEEEtran}
\ifCLASSOPTIONcompsoc
  \usepackage[nocompress]{cite}
\else
  \usepackage{cite}
\fi
\ifCLASSINFOpdf
\else
\fi
\usepackage{subfig}

\usepackage{comment}
\usepackage{tikz}
\usepackage{graphicx}
\usepackage{color}
\usepackage{tabularx}
\usepackage{bchart}

\usepackage{wrapfig}
\usepackage{multirow}
\usepackage{algorithm} 
\usepackage{algpseudocode} 

\begin{document}
%
\title{Hydra: Hybrid Server Power Model}
\author{Nigel Bernard$^1$,
    Hoa Nguyen$^1$,
    Aman Chandan$^2$,  
    Savyasachi Jagdeeshan$^2$,\\ 
    Namdev Prabhugaonkar$^2$, 
    Rutuja Shah$^2$,
    Hyeran Jeon$^1$\\
    \small{$^1$University of California Merced, 
    $^2$San Jose State University}}


%



\markboth{IEEE Computer Architecture Letter}%
{Chandan \MakeLowercase{\textit{et al.}}: DeepPower: Deep Learning Accelerated Server Power Prediction}

\IEEEtitleabstractindextext{
\begin{abstract}
With the growing complexity of big data workloads that require abundant data and computation, data centers consume a tremendous amount of power daily. In an effort to minimize data center power consumption, several studies developed power models that can be used for job scheduling either reducing the number of active servers or balancing workloads across servers at their peak energy efficiency points. Due to increasing software and hardware heterogeneity, we observed that there is no single power model that works the best for all server conditions. Some complicated machine learning models themselves incur performance and power overheads and hence it is not desirable to use them frequently. There are no power models that consider containerized workload execution. In this paper, we propose a hybrid server power model, \textit{Hydra}, that considers both prediction accuracy and performance overhead. Hydra dynamically chooses the best power model for the given server conditions. Compared with state-of-the-art solutions, Hydra outperforms across all compute-intensity levels on heterogeneous servers. 
\end{abstract}

}

\maketitle
%
\IEEEdisplaynontitleabstractindextext
\IEEEpeerreviewmaketitle

\section{Introduction}
In 2020, data centers in the USA consumed 91 billion electricity units (kW/h) yielding \$13 billion per year for electricity bills in the business sector. Some of the world’s largest data centers can each contain many tens to thousands of computing devices and require more than 100 megawatts of power capacity, which is enough to power around 80,000 U.S. households. The significance of maximum energy efficiency in data centers cannot be overstated.

To reduce such an excessive data center power usage, several solutions have been proposed. Some studies demonstrated dynamic load control algorithms that either reduces the number of active servers by consolidating workloads to fewer servers~\cite{verma_atc09} or balances the loads across servers such that all servers can run under the peak energy efficiency point~\cite{daniel_isca16}. Most of these studies used either CPU utilization or linear or polynomial power prediction models to find the migration triggering point. The server power prediction models use CPU utilization as the main or even the only parameter as CPU utilization is known to be highly correlated with overall system power consumption. 
However, under increasing heterogeneity in both software and hardware~\cite{heterogeneous_statistics}, there is no single model that rules all possible server configurations. For example, unlike many conventional analytical models assumed, CPU is no longer the only significant power consumer in server systems. In big data era, heavy memory usage is an inevitable computing pattern. Likely, accelerators and virtualized computing environments have notable impacts towards server power consumption. 
Therefore, the effectiveness of CPU-utilization-based server power management is questionable. 

\begin{figure}
\centering
  \includegraphics[width=0.30\textwidth]{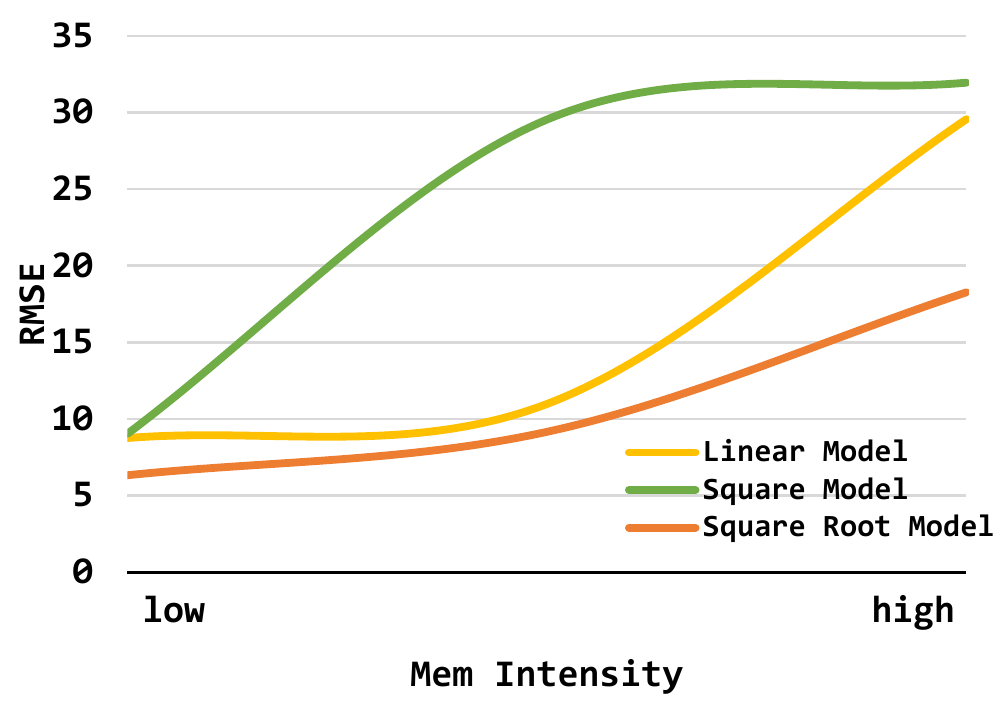}
\caption{Prediction Errors w.r.t. Memory Intensity}
\label{fig.motivation}\vspace{-15pt}
\end{figure}

Figure~\ref{fig.motivation} shows the prediction accuracy of three CPU-utilization-based power models with regarding to memory intensity of applications in root mean square error (RMSE). The models are brought from a power modeling study in cloud simulation frameworks~\cite{powermodel}. The models use the linear, the square, and the square root of CPU utilization to the following model for power prediction, where $\alpha$ value is 1, 2, 1/2 for linear, square, and square root models, respectively. 

$Server\ Power = P_{min} + (P_{max} - P_{min})\times Util_{CPU}^{\alpha}$

\noindent Across all the models, the error rate raises almost exponentially as the memory intensity increases, where the averaged power prediction error for the high memory intensive applications is greater than 30 Watts, which is not negligible.

To accommodate the hardware heterogeneity, some studies included non-compute components such as memory, storage, and network for the power modeling~\cite{roy_itcs13, marius_sigmetrics13, ge_ipdps09}. However, it is not easy to include new parameters to existing analytical models and find the optimal weight values, which requires significant engineering efforts. 
Furthermore, adding more parameters to the existing analytical models is not a scalable solution given the ever-increasing system heterogeneity. For a hassle-free model development, some recent studies adopted machine learning algorithms~\cite{lstm_powerprediction, lstm_powerprediction2}. To reflect non-linear impact of various system parameters towards server power consumption, some other researchers used deep learning algorithms~\cite{dnn-powerprediction, dnn-powerprediction2, dnn-powerprediction3}. These studies used tens of system parameters and the corresponding server power values collected for days even months of time to train a power prediction model. These models show superior accuracy than many conventional models. However, given the inherent compute intensity of machine learning algorithms, machine learning-based power models themselves may cause  performance and power overhead. Our experiments show that a deep learning model takes at least 25$\times$ longer latency and consumes 3 $\sim$ 50$\times$ more power than a polynomial model. Therefore, the usage of those compute-heavy power models should be carefully determined so that the models can be used only when necessary. 

To mitigate the aforementioned limitations of existing power models, we propose a novel hybrid server power model namely \textit{Hydra} that tackles both the accuracy deficiencies of conventional analytical models and the computational overhead of machine learning-based models. Hydra incorporates a lightweight server workload classifier that selects the best power model that considers both performance overhead and prediction accuracy. 
In our experiments conducted in a small-scale data center that consists of heterogeneous servers, Hydra showed the superior prediction accuracy and less performance overhead for various workload sets consisting of different compute and memory intensity compared to several state-of-the-art power models. 
\section{Related Work and Motivation}

\subsection{Machine Learning for Server Power Prediction}
Shen et al.~\cite{lstm_powerprediction} used recurrent neural network (RNN) for the server power prediction, especially by using long short-term memory (LSTM) algorithm. They used LSTM for future power of different time granularity: immediate (next second) and further future (next 30 seconds). Li et al.~\cite{dnn-powerprediction2} used recursive auto-encoder (RAE) to predict server power by using power history data, per-server system counters, and total power draw of a data center. 
Lin et al.~\cite{dnn-powerprediction3} evaluated four artificial neural network (ANN) power models including time window back propagation neural network (TWBPNN) that considers both power history and system parameters. They used 16 system parameters. 
Most of the studies did not clearly reveal the correlation between the parameters and server power. Also, all of them used bare-metal workload executions that are not common in data centers today. Unlike these studies, we provide a clear correlation between system parameters and the server power consumption, which enables to extract the minimum number of parameters and we consider containerized execution, which is more realistic to the current data center configuration. 

\subsection{Why Do We Need Another Power Model?}

\begin{figure}
\centering
  \subfloat[][Un-containerized Execution ]{\includegraphics[width=0.4\textwidth]{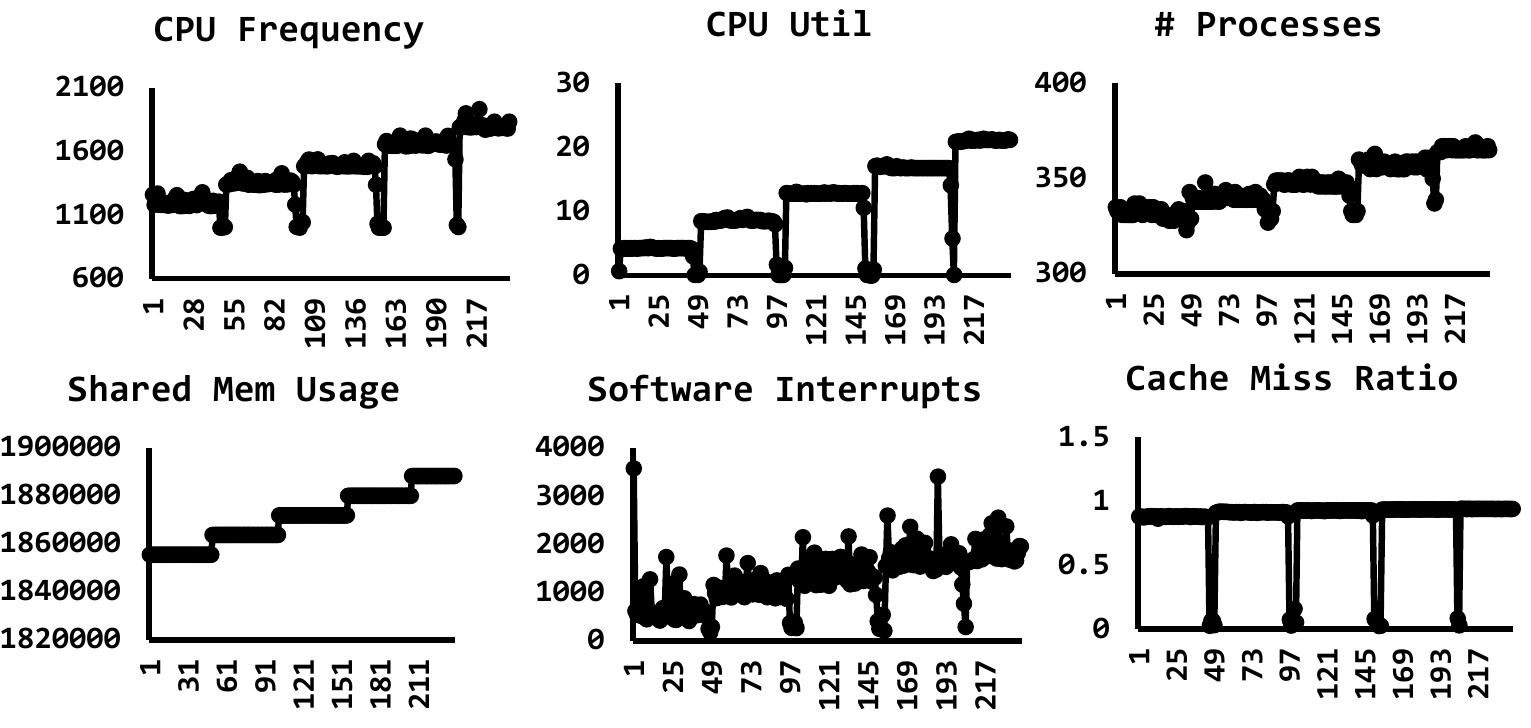}}\\
  \subfloat[][Containerized Execution]{\includegraphics[width=0.4\textwidth]{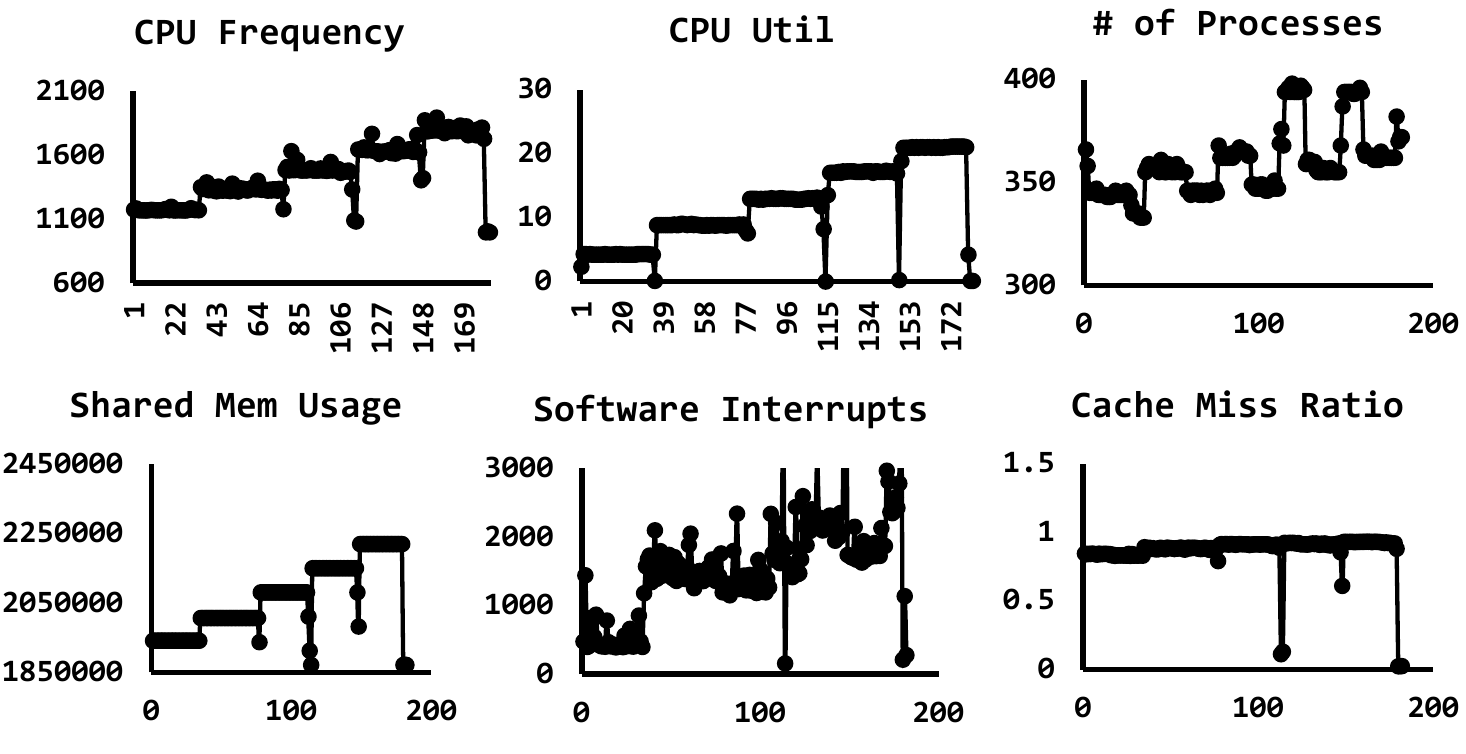}}
\caption{System Param Correlation with/without Container while increasing the number of \textit{streamcluster from PARSEC} instances from 1 to 5 (Y-axis : parameter, x-axis : time). } \label{fig.container}\vspace{-15pt}
\end{figure}

Though server power modeling has been extensively studied, due to the increasing heterogeneity in both hardware and software, the effectiveness of existing models would be diminished. For example, we already showed with Figure~\ref{fig.motivation} that CPU utilization-based server power models derive high error rate as workloads' memory intensity increases. 
Virtualized workload executions with containers also drops accuracy of existing power models. Figure~\ref{fig.container} shows the results when five instances of an application are executed with a fixed interval (a) without container and (b) with container. Some parameters show clear impact of each application instance with staircase-patterned increasing graph (i.e., CPU freq. in both containerized and uncontainerized executions). But, for some other parameters such as software interrupts and cache miss ratio, it is hard to tell when new application instance is invoked, due to shared helper processes used by containers, which lead to a much loose correlation between individual workloads and server power consumption.

\begin{figure}
\centering
  \subfloat[][Intel Xeon Cascade Lake-SP Silver-4214 ]{\includegraphics[width=0.50\textwidth]{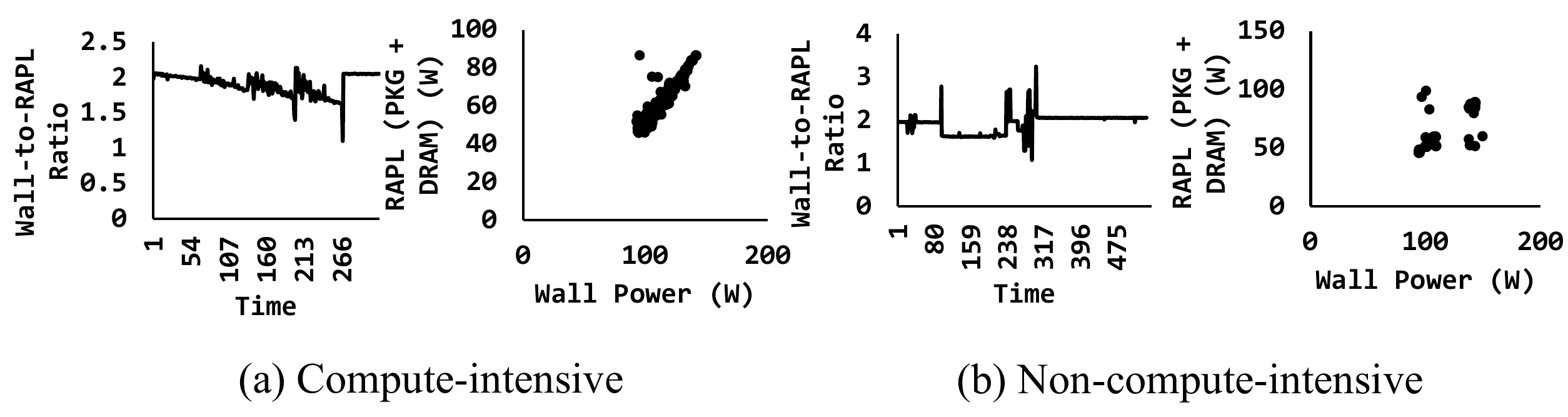}}\\ \vspace{-10pt}
  \subfloat[][Intel Xeon Sandy Bridge-EP E5-2620 ]{\includegraphics[width=0.50\textwidth]{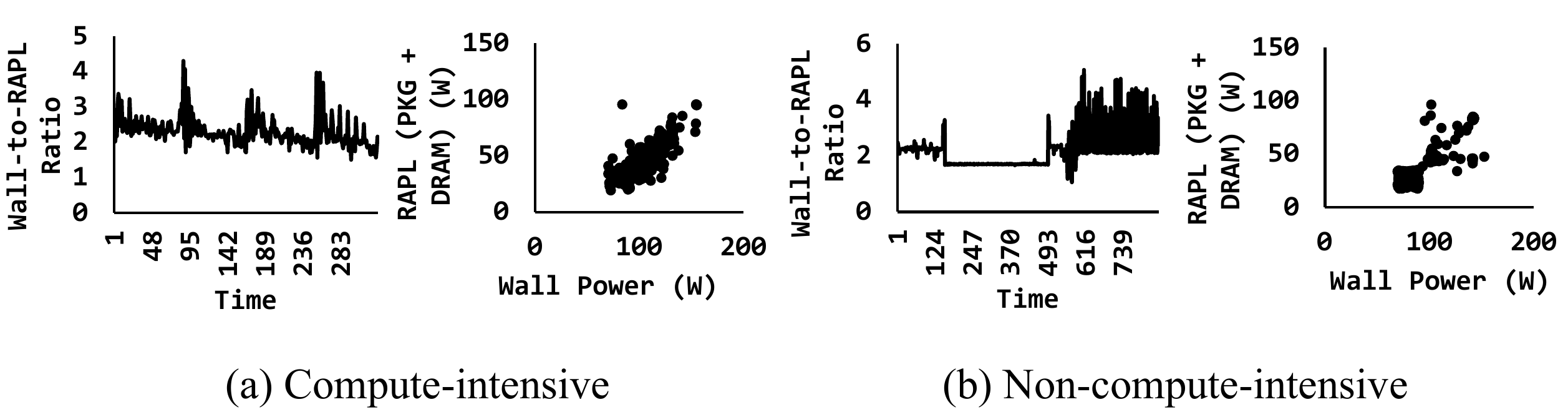}}\\
\caption{Wall Power vs. RAPL measurements on Two CPU Generations} \label{fig.rapl}\vspace{-15pt}
\end{figure}

Some studies used processor-provided power profilers instead of power models, such as Intel's RAPL. RAPL power values are known to be highly accurate in newer CPUs. However, according to our experiments, RAPL data does not reflect server power consumption because it only shows the processor power. Figure~\ref{fig.rapl} shows the ratio of server power to RAPL measurements while running different type workloads on two different Intel CPUs. In each pair of graphs, the left shows wall power-to-RAPL ratio during the workload execution and the right shows the relation between the wall power and RAPL. If RAPL can substitute server power models, any of the following two conditions should be satisfied: 1) the left graphs show a consistent value, 2) the right graphs show a clear trend line. However, none of the results show such relation. The irregularity is more severe with non-compute workloads and older CPU models. Though some studies explored ways to predict server power model by using RAPL result as a parameter~\cite{rapl_powermodel}, given the high irregularity between the server and RAPL power, we believe it is not effective to use RAPL to predict server power model. \textbf{Therefore, we believe a new server power model is needed that can easily scale with increasing heterogeneity.}

\section{Hydra, The Server Power Model}\label{sec.hydra}

The proposed power model, Hydra, considers not only accuracy but also performance overhead by selecting an optimal model for the given server condition rather than sticking to one certain model every time. To use the optimal model, Hydra implements 1) a \textbf{power model selector} that monitors the current system statistics of the target server, classifies the server condition to be mapped to one (or multiple) candidate power models that show good accuracy in similar conditions, and chooses one power model that has the lightest performance overhead, and 2) a \textbf{power predictor} that estimates the server power consumption by feeding the system statistics to the selected power model. 

\begin{figure}
\centering
  \subfloat[][Compute-intensive Workloads (Concurrent executions of CPU Stress Tests of Stress-NG benchmark suite) ]{\includegraphics[width=0.4\textwidth]{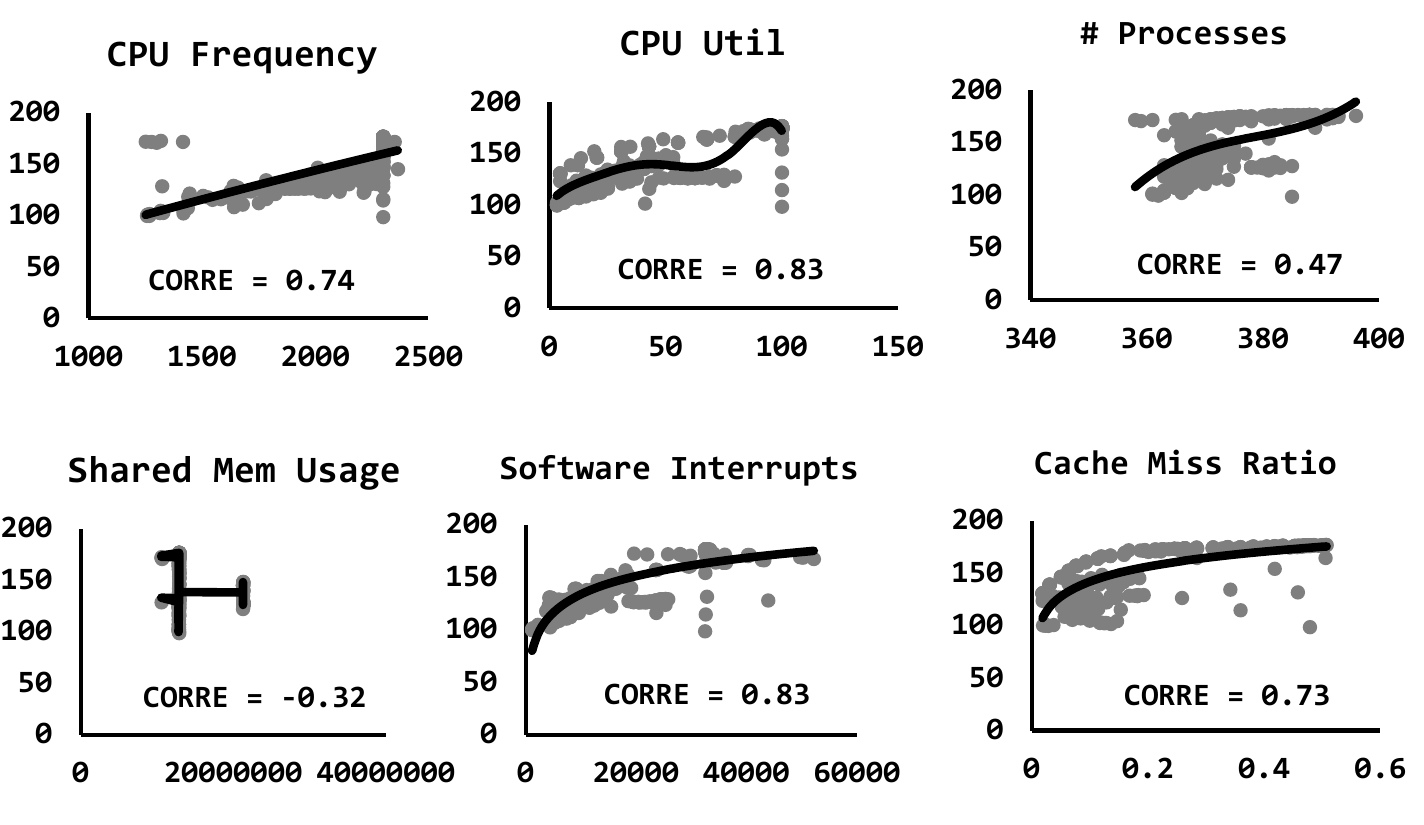}}\\ \vspace{-10pt}
  \subfloat[][Non-compute-intensive Workloads (Concurrent executions of canneal, streamcluster, swaptions, and raytrace of PARSEC benchmark suite)]{\includegraphics[width=0.4\textwidth]{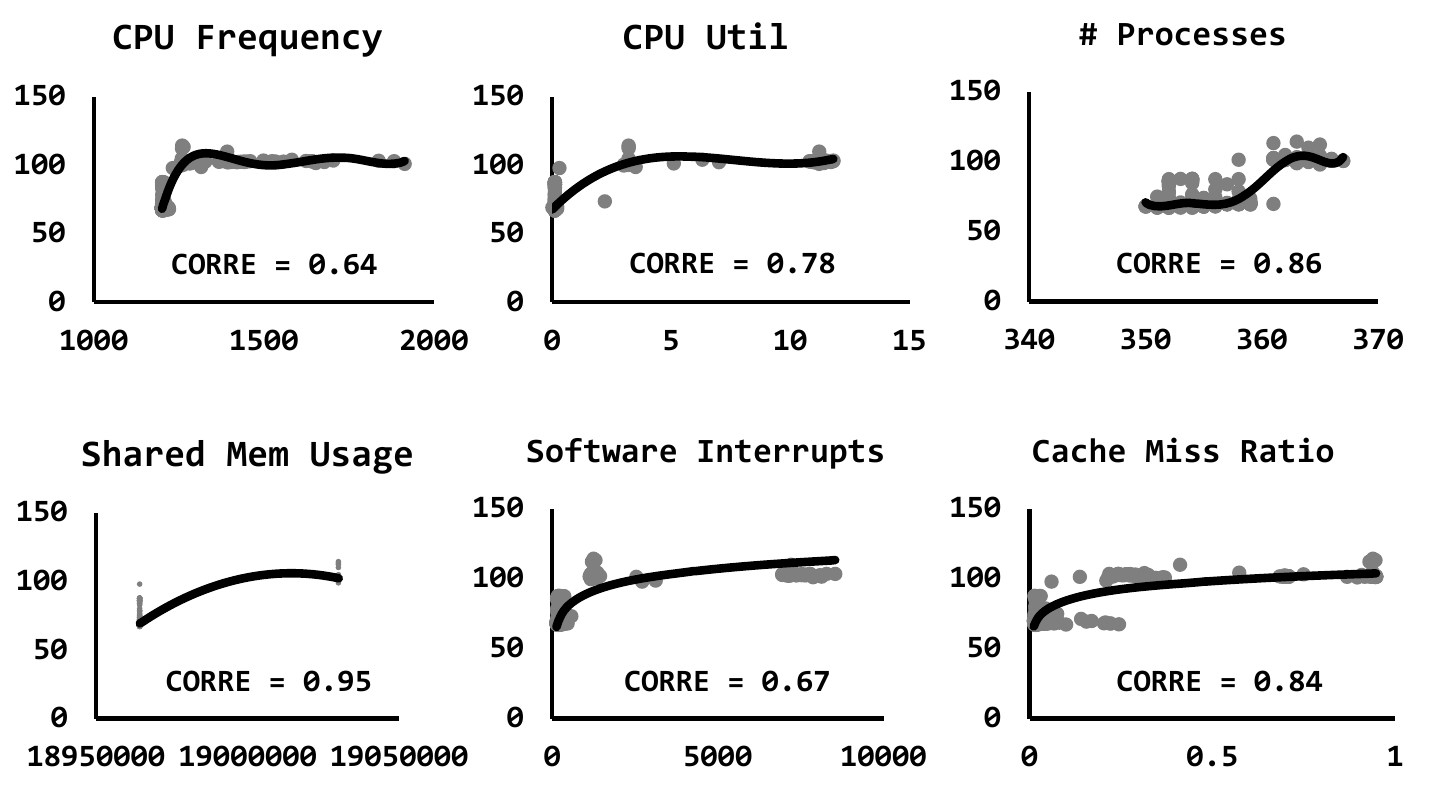}}
\caption{System Parameters (x-axis) Correlation with Server Power (y-axis in Watt): Gray dots - real measurements. Black lines - trend lines. CORRE - Pearson's correlation.}\vspace{-15pt} \label{fig.params}
\end{figure}

\textbf{Power Model Selector: }
Though more complex models may show better accuracy regardless the system conditions, as we also consider performance overhead, it is not that simple to find the best power model. For example, a lightweight CPU-utilization-based analytical model may derive good-enough accuracy when the server runs mostly compute-intensive workloads, while a complex machine learning model may be required when the server runs a combination of various types of workloads at the cost of performance. Likely, for machine learning models, different algorithms show diverse accuracy for different server conditions. For example, RNN would be better for the servers that show regular patterns over time while fully-connected deep neural network (DNN) may be better for those that cannot clearly expect workload types and scheduling times. Therefore, we propose to develop a power model selector that identifies a power model that will show the best accuracy for the given server condition with the lightest performance overhead. As the selector itself should not incur extra performance overhead, we use a lightweight regression model. While different algorithms may work better depending on server power pattern and candidate power models, we chose to use a random forest model that is trained with over 2K workload datasets (containing system statistics and power measurements for various compute intensities) as it showed superior selection accuracy than other tested algorithms such as decision tree and k-nearest neighbor models in our experiments. The random forest model chooses one of the candidate power models based on current system statistics. 
Note that system statistics themselves (e.g., CPU utilization and memory usage) are not sufficient to select the best power model because in many cases server conditions are not clearly distinguishable with one or two system stats and individual models show diverse accuracy for different server conditions.


\textbf{Server Power Predictor: }
Though types of power models can vary, we used two extreme models, an analytical model for the minimal overhead and a machine learning model for the best accuracy. For the analytical model, we examined various existing models and chose the linear model because the model showed comparable accuracy with the best performance as shown in Figure~\ref{fig.motivation}.
For the machine learning model, it is important to carefully identify an algorithm that well represents the server power from system parameters.  
To understand the diverging contributions of individual system components towards the server power, we examined 28 system statistics and the server power consumption while running two distinctive sets of workloads. The system statistics were collected by using two Linux commands, \textit{perf} and \textit{psutil}. 
Figure~\ref{fig.params}(a) shows the statistics of representative parameters while running compute-intensive workloads and Figure~\ref{fig.params}(b) shows the same statistics of non-compute-intensive workloads. 
For the compute-intensive workloads, we used Stress-NG benchmark suite,  
while, for the non-compute-intensive workloads, we ran four applications of PARSEC benchmark suite concurrently that have high misses per kilo instructions (MPKI). All workloads were containerized to make the configuration realistic to server systems. In each graph, y-axis is the server power consumption and x-axis is the scale of the target system parameter. The numbers in the graph are Pearson's correlation coefficient between the parameter and the server power consumption. The gray dots are real measurements and the black lines show trend line of the measurements. 

As can be seen, the relations show significantly different shapes depending on the workloads type. For the compute-intensive workload, the power shows almost linear relation with CPU frequency, while that of non-compute-intensive workloads is represented the best with a five-degree polynomial model. Likely, non-compute-intensive workloads show high correlation with shared memory usage (correlation is over 95\%), while compute-intensive workloads show almost no correlation between the shared memory usage of the server power consumption. While all these parameters show strong correlation (over 70\%) with server power in either of the workload types, there are no two parameters that show exactly matching behaviors. To accommodate such diverse contributions of all these parameters across different workload types, we believe a more scalable solution is needed. To reflect non-linear relations of these tens of parameters, we propose to use a fully-connected DNN model.

\begin{table}[]
\centering
\scalebox{0.8}{
\centering
\begin{tabular}{|l|l|}
\hline
{\textbf{Parameter}}       & {\textbf{Description}}    \\ \hline
{\textbf{CPU Frequency}}       & {Frequency of the CPU}     \\ 
{\textbf{User Time}}       & {Time spent by the CPU in user space}    \\ 
{\textbf{CPU Util}}       & {Percentage of CPU used}    \\ 
{\textbf{Interrupts}}       & {Number of Interrupts}    \\ 
{\textbf{S/W Interrupts}}       & {Number of Software Interrupts}    \\ 
{\textbf{Processes}}       & {Number of processes in the system}    \\ 
{\textbf{Cache Miss Ratio}}       & {Cache miss ratio}    \\ 
{\textbf{Virtual Mem Usage}}       & {Percentage of virtual memory used}    \\ 
{\textbf{System Calls}} & {Number of System Calls}    \\ 
{\textbf{Instructions}} & {Number of instructions executed}    \\ 
{\textbf{Shared Mem Usage}}       & {Number of bytes used in shared memory}    \\ \hline
\end{tabular}}
\caption{Selected System Stats Used for Power Model}\label{tab:dnn_parameters}\vspace{-15pt}
\end{table}

\textit{Model Structure: }
We evaluated various fully-connected DNN structures with Talos tool~\cite{talos} and designed the final version to have six hidden layers with 16, 32, 64, 32, 16, and 8 neurons per layer, which derived the best accuracy.

\textit{Model Inputs: }
From the experiments, we observed that not all 28 system parameters show strong correlation with the server power consumption. Due to weak correlation, some parameters only degrade model accuracy. Thus, we identified the parameters that show strong correlation with the power consumption. We selected 11 parameters that show higher than 70\% correlation as listed in Table~\ref{tab:dnn_parameters}.

\textit{Model Outputs: }
We observed that servers exhibit different idle and peak power consumption due to different size of computing resources, energy-proportionality of processors, and so on. To tackle the server heterogeneity, we propose to predict power scale factors rather than absolute power values. 
The power scale factor indicates the relative power level out of the maximum power range (between idle power and peak power), which is calculated as $(Power_{current}\ -\ Power_{idle})/(Power_{max}\ -\ Power_{idle})$. Once the power model predicts the power scale factor, we compute the predicted absolute power from it based on individual servers' idle and max power values.

\section{Evaluation}\label{sec.eval}

We used 13 workloads for our evaluation and data collection from PARSEC and SPLASH benchmark suites. We also used stress micro-benchmarks from Linux Stress-NG to evaluate on various compute intensity conditions. We tested the models on two types of servers each uses different Intel Xeon CPU generation: Cascade Lake-SP (32 cores) with 384KB L1, 12 MB L2, 16.5 MB L3 and Sandy Bridge-EP (24 cores) with 32KB L1, 256KB L2, 20 MB L3. 
We compared the performance of our approach with RAE~\cite{dnn-powerprediction2} and TWBPNN~\cite{dnn-powerprediction3}, as well as a simple LSTM-based RNN model that uses past 30 power values for prediction. All models were trained with our system parameters. We tried to match the parameters as close to those in original papers. 

\begin{figure}
\centering
  \includegraphics[width=0.4\textwidth]{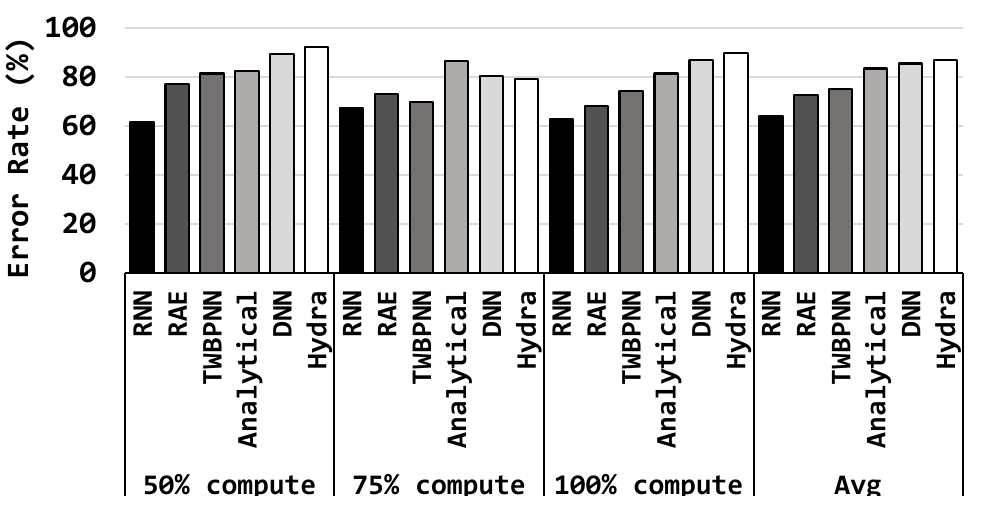}
  \caption{Prediction Accuracy w.r.t. Compute Intensity} \label{fig.accuracy}\vspace{-15pt}
\end{figure}

The performance and power overheads of individual models are summarized in Table~\ref{tab:model_overheads}. The prediction accuracy for different compute intensities is plotted in Figure~\ref{fig.accuracy}. The overhead and accuracy are measured while running the workloads with long-tail job distribution with mixed compute intensities. 
Hydra's overhead is the averaged overhead that includes the execution of model selector and the selected power model. RNN showed the longest execution time due to its sequential execution model while the analytical model showed almost negligible performance overhead. Hydra showed less performance and power overhead than most of the other machine learning models except for RAE. But, the prediction accuracy of RAE was over 10\% worse than Hydra because RAE does not consider server heterogeneity, while Hydra is trained with power scale factors with more system parameters. Hydra showed superior overall accuracy thanks to the dynamic model selection. 

\begin{table}[]
\centering
\begin{tabular}{|c|c|c|}
\hline
Model          & Latency (ms) & Power (Watt) \\ \hline\hline
Analytical & $$\math{<}$ 0.001 &    $$\math{<}$ 0.2       \\ \hline 
DNN        & 0.025          & 9.52          \\ \hline 
RNN        & 1.003          & 11.55         \\ \hline 
Hydra      & 0.07           & 5.64          \\ \hline
RAE~\cite{dnn-powerprediction2}        & 0.039          & 0.61          \\ \hline 
TWBPNN~\cite{dnn-powerprediction3} & 0.092          & 9.42          \\ \hline
\end{tabular}
\caption{Per-Model Overheads: Idle power of the tested server is 237.58 Watt}\label{tab:model_overheads}\vspace{-15pt}
\end{table}

\vspace{-5pt}
\section{Discussion}
Due to limited space, we show results of Hydra that uses only two candidate power models. However, Hydra pursues higher scalability and flexibility. There are no limitations of the number of candidate power models. The extended Hydra will be able to support various emerging workloads and hardware components that were not well supported with earlier power models. By selecting the lightest power model at run time, Hydra can be integrated to OS schedulers such that data centers' overall power consumption can be reduced through power-aware job assignment without expensive hardware power meters integrated to every server. The power models also can be updated at runtime by integrating a power meter per server type, collecting system statistics and power measurements, and invoking transfer-learning thread regularly. We will discuss these ideas in the extended full-length paper. 
\vspace{-5pt}
\section{Conclusion}
Due to increasing system heterogeneity, there is no one outperforming power model. Our proposed hybrid prediction model dynamically chooses the best power model with the consideration of prediction accuracy and model's performance overhead. Our simple combination of an analytical and a DNN model showed superior performance and accuracy for various compute intensity. 

\bibliographystyle{IEEEtran}
\bibliography{bibliograph, nigelbib}

\end{document}